\documentclass[a4paper,11pt]{iopart}
\usepackage{graphicx}
\usepackage{cite}
\usepackage{color}
\usepackage{amsfonts}
\usepackage{srcltx}
\begin{document}
\title{Characteristics of a random walk on a self-inflating support}
\author{Lukas Kades$^1$, Manuel Schrauth$^2$, Maximilian Schneider$^2$ and Haye Hinrichsen$^2$}
\address{Universit\"at Heidelberg, Fakult\"at f\"ur Physik und Astronomie, Im Neuenheimer Feld, 69120 Heidelberg, Germany}
\address{Universit\"at W\"urzburg, Fakult\"at f\"ur Physik und Astronomie, Am Hubland, \\ 97074 W\"urzburg, Germany}

\begin{abstract}
Self-similar dynamical processes are characterized by a growing length scale $\xi$ which increases with time as $\xi \sim t^{1/z}$, where $z$ is the dynamical exponent. The best known example is a simple random walk with $z=2$. Usually such processes are assumed to take place on a static background. In this paper we address the question what changes if the background itself evolves dynamically. As an example we consider a random walk on an isotropically and homogeneously inflating space. For an exponentially fast expansion it turns out that the self-similar properties of the random walk are destroyed. For an inflation with power-law characteristics, however, self-similarity is preserved provided that the exponent controlling the growth is small enough. The resulting probability distribution is analyzed in terms of cumulant ratios. Moreover, the dynamical exponent $z$ is found to change continuously with the control exponent.
\end{abstract}

\def\d{{\rm d}}
\def\0{\emptyset}
\def\comment#1{\color{red}[\textbf{comment: #1}]\color{black}}
\def\mark#1{\color{red}#1 \color{black}}
\newcommand{\eqref}[1]{(\ref{#1})}
\newcommand{\Eqref}[1]{Eq.~\eqref{#1}}

\pagestyle{plain}

\section{Introduction}

One of the most important cornerstones for our understanding of critical phenomena is the concept of self-similarity and scale invariance~\cite{Cardy1996}. These terms describe a situation in which the physical state of a system is in some sense invariant under a change of scale of the supporting geometry. Scale-invariant properties are often found to be universal, i.e. they depend only on the symmetries of the system but not on the specific microscopic realization. 

On a lattice, a scale transformation can be carried out by coarse-graining the elementary degrees of freedom. One of the best known examples is Wilson's block-spin renormalization of the Ising model, where several spins are grouped into blocks~\cite{Fisher,Wilson,Kadanoff}. If the Hamiltonian in terms of these block variables has the same form as the original one (possibly with different parameters), this procedure maps the system onto itself, forming a group of renormalization transformations. The analysis of the renormalization group (RG) flow and the corresponding fixed points reveals essential information that characterizes the critical behavior.

In most studies, scale invariance is used as a mathematical tool for the analysis of critical phenomena. The aim of the present study is to view scale transformations from a different perspective, namely, as part of the physical process itself. More specifically, we will consider time-dependent processes in which infinitesimal scale transformations are continuously carried out as part of the dynamics. This would correspond to a self-inflating (or self-deflating) supporting geometry on which the process takes place. If the process itself is scale-invariant, it would be interesting to study how it responds to the ongoing change of scale of the underlying support. As a possible motivation, such self-inflating scale-free processes may be viewed as toy models for critical phenomena taking place in an expanding universe.

\begin{figure}
\centering\includegraphics[width=130mm]{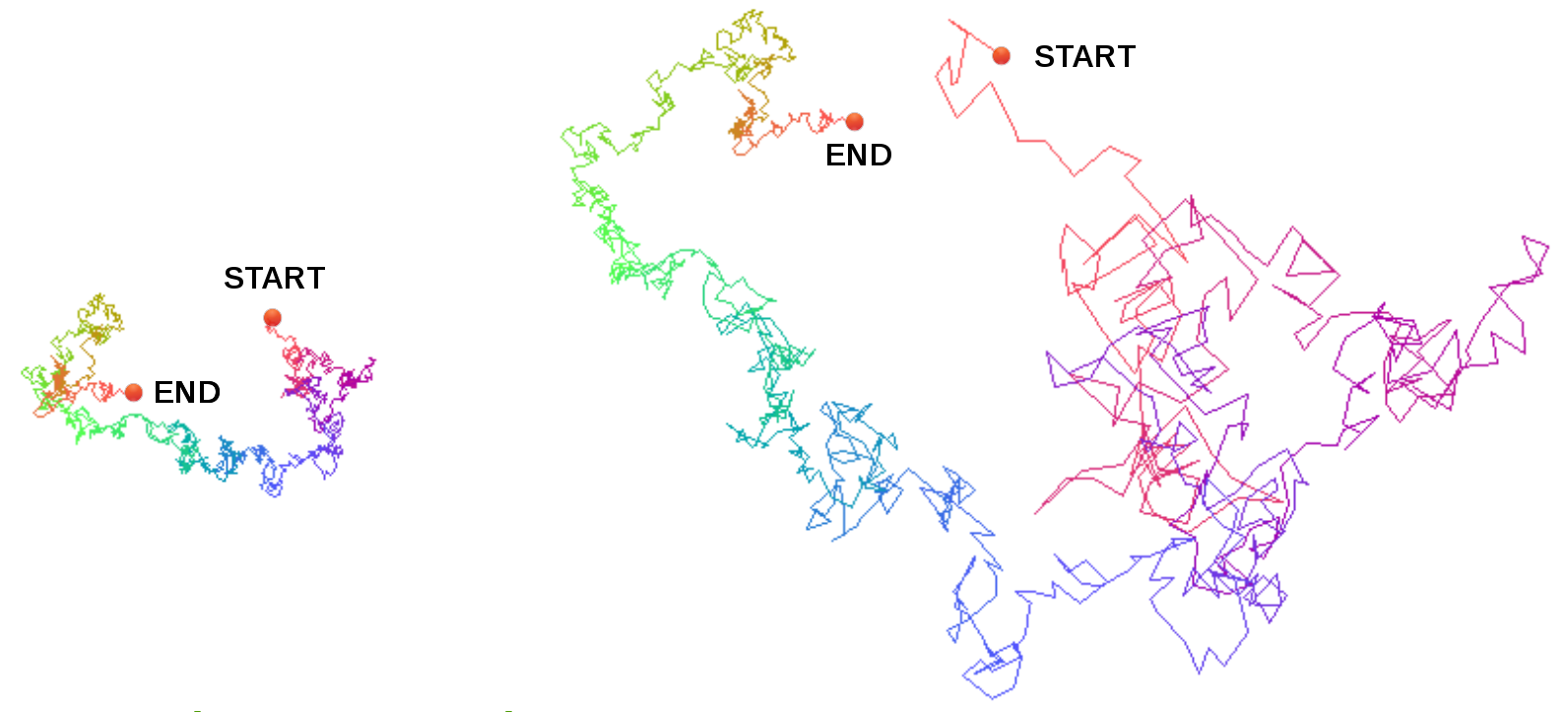}
\label{fig:rw}
\caption{\textbf{Random walk on a self-inflating support.} Left: Ordinary random walk in a static two-dimensional plane. The changing color scale indicates the advance of time. Right: The corresponding random walk with exactly the same sequence of microscopic displacements taking place on a self-inflating two-dimensional plane (see text). Shown is the situation at the end of the simulation. Consequently, earlier steps have undergone a stronger expansion compared to later ones.}
\end{figure}

As a simple example let us consider a random walk in two dimensions, as shown in Fig.~\ref{fig:rw}. The left side of the figure shows an ordinary random walk on a static background consisting of $1000$ statistically independent displacements in random directions. The right side shows the same random walk with an identical sequence of steps on a self-inflating background, dilating the underlying space during each step by the factor $1.002$. Illustratively stated, this could be the random path of an ant on an inflating balloon. 

It is important to note that the random walk on the right side consists of exactly the same sequence of microscopic displacements as the one shown on the left side. However, early moves are significantly stretched by the ongoing dilatation, leading in total to a completely different trajectory of the random walker. Obviously, a random walker on a self-inflating support moves faster away from the origin than in the static case. Are the resulting distances still normally distributed? If so, how do they grow with time? 

The purpose of this paper is to investigate these questions in detail. To this end we first discuss the phenomenological scaling properties of a self-similar stochastic process on a homogeneously inflating background. In Sect.~3, we recall how random walks can be analyzed in terms of cumulants. We then analyze a random walk on an expanding background both numerically and by analytical arguments. Here, we consider two particular types of inflation, namely, exponential and power-law driven expansion. The paper ends with concluding remarks.

\section{Phenomenological scaling on an expanding background}

Consider a physical Euclidean space which is expanding homogeneously and isotropically by itself. Choosing one of its points as the origin, any other point at location $\vec x(t)$ will move away from the origin. More specifically, the corresponding position vector $ \vec{x}(t) $ will grow according to the differential equation
\begin{equation}
\label{eq:selfinflation}
\frac{\d}{\d t}{\vec x}(t) = H(t)\,\vec x (t)\,,
\end{equation}
where the scalar quantity $H(t)$ is the (possibly time-dependent) expansion rate. Alluding to cosmology, this function will be referred to as the Hubble parameter. For a given Hubble parameter, we can define the dimensionless scale factor $a(t)$. This scale factor evolves by the differential equation
\begin{equation}
H(t)=\frac{\dot a(t)}{a(t)}
\end{equation}
with the solution
\begin{equation}
a(t) \;=\; a(t_0) \,\exp \Bigl[ \int_{t_0}^t \d t'\, H(t') \Bigr].
\label{eq:scalefactor}
\end{equation}
The scale factor $a(t)$ tells us by which factor the length scales have been inflated with respect to a certain reference time $t_0$. Since we want to start from a non-stretched system, we set $ a(t_0)=1 $.

Let us now consider a scale-free dynamical critical phenomenon on this self-inflating space. The scaling hypothesis states that the physical state of such a system is invariant under rescaling of space and time by $\vec x \to a \vec x$ and $t \to a^z t$, where $a>0$ is a scale factor and  $z=\nu_\parallel/\nu_\perp$ is the so-called dynamical exponent. In addition, for a random walk the probability density $p(\vec x,t)$ has to be rescaled by $p \to a^{-1}p$ in order to preserve normalization.~\footnote{More generally, for an arbitrary nonequilibrium critical phenomenon, the order parameter $\rho$ has to be rescaled by $\rho\to a^{-\beta/\nu_\perp}\rho$, where $\beta$ is the critical exponent associated with the order parameter.}

Assuming that the random walk is still self-similar even on a self-inflating support, we expect it to be invariant under the scale transformation
\begin{equation}
\vec x \to a(t)\, \vec x\,, \qquad t \to a(t)^z t \,, \qquad p \to a(t)^{-1} p\,.
\label{eq:scaling-transformation}
\end{equation}
As a result, the probability distribution of the random walker should still be given by a Gaussian stretched by the ongoing expansion. This could be used to 'accelerate' the buildup of the correlation length in such a process. However, it is intuitively clear that we cannot accelerate this buildup as much as we like because the expansion does not only stretch the large-scale structure but also the small-scale deviations caused by lattice effects. It could even be that the underlying geometry expands faster than the correlation length of the random walk itself, freezing its current structure on large scales.\footnote{This may have happened, for example, with quantum fluctuations during the inflationary period of the early Universe, which now manifest themselves as frozen irregularities on large scales in the microwave background~\cite{Liddle2009}.} In fact, as we see below, this will depend on the expansion characteristics, i.e., on the choice of the function $a(t)$.

\section{Characterizing a random walk by its cumulants}

\subsection*{Cumulants as a measure of non-Gaussianity}
Does a random walk on a self-inflating background still obey the central limit theorem? To answer this question we study the \textit{cumulants} of the corresponding probability densities. These quantities were first introduced by Thiele (1889) and have been given the name \textit{cumulants} by Fisher and Wishart (1932). A comprehensive account of their derivation and properties can be found in many standard textbooks on statistics, such as e.g., Kendall (1969) and can be summarized as follows:

Recall that for a random variable $X$ with real values $x \in \mathbb R$ distributed according to the probability density $P(x)$ the cumulants are defined by
\begin{equation}
\label{eq:cumulants}
\kappa_n \;:=\; (-i\partial_k)^n \ln  \varphi(k) \Bigr|_{k \to 0}\,,
\end{equation}
where 
\begin{equation}
\varphi(k) \;=\; \int_{-\infty}^{\infty} \d x\, e^{i k x} P(x) 
\end{equation}
is the so-called \emph{characteristic function}. The cumulants are combinations of moments which are chosen in such a way that for a normal distribution $P(x)=\frac{1}{\sigma\sqrt{2 \pi}} e^{-(x-\mu)^2/2\sigma^2}$ all cumulants except for the mean $\kappa_1=\mu$ and the variance $\kappa_2=\sigma^2$ are equal to zero. This property is unique to the normal distribution. Therefore, in order to check if a certain distribution is Gaussian it is sufficient to show that all cumulants $\kappa_n$ for $n>2$ vanish identically. To this end, it is convenient to study the dimensionless cumulant ratios 
\begin{equation}
	R_n :=\frac{\kappa_n}{\kappa_2^{n/2}} \quad\quad\mbox{for}\quad n\geq2
	\label{eq:cumulant_ratio_definition}
\end{equation}
as a measure for the non-Gaussianity of a distribution. The latter are especially useful in cases where the bare cumulants of a sum of random variables diverge, since they are standardized by the width of the underlying distribution.

The cumulants feature some useful properties. At first, it can easily be shown from Eq.~\eqref{eq:cumulants} that in the case of a sum of two statistically independent random variables $ X_1+X_2 $, where the corresponding characteristic functions multiply according to the convolution theorem, the cumulants simply add up, i.e.
\begin{equation}
	 \kappa_n[X_1+X_2] = \kappa_n[X_1] +\kappa_n[X_2] \qquad n \in \mathbb{N}.
\end{equation}
Secondly, the $ n $-th cumulant is a homogeneous function of degree $ n $, meaning that
\begin{equation}
	\kappa_n[cX]=c^n \kappa_n[X]
\end{equation}
for all $ c>0 $. Both relations will be used in what follows.

\subsection*{Cumulants of an ordinary one-dimensional random walk on a static background}
As mentioned before, a random walk will be asymptotically Gaussian if all higher cumulants $ \kappa_3,\kappa_4,\ldots $ (or cumulant ratios $R_3,R_4,\ldots$) tend to zero as the number of steps increases. Let us first verify this criterion in the case of an ordinary (non-inflating) symmetric one-dimensional random walk
\begin{equation}
\label{eq:OrdinaryRW}
x_{j+1} := x_j + r_j \qquad \Rightarrow \qquad x_N = \sum_{j=0}^{N-1} r_j
\end{equation}
with the initial condition $x_0=0$ and statistically independent random displacements $r_j \in \mathbb R$. For simplicity let us assume that all displacement are drawn from a flat distribution between $-1$ and $1$, i.e. their probability density is given by
\begin{equation}
P(x) \;=\; \mathcal{U}_{(-1,1)}(x)\,=\,
\left\{\begin{array}{cl}
\frac{1}{2} & \mbox{ for }\; -1< x < 1, \\
0 & \mbox{ otherwise. }
\end{array}\right.\,
\label{eq:uniform_distribution}
\end{equation}
With the corresponding characteristic function  $\varphi(k) = \frac{\sin k}{k}$ it is straight-forward to calculate the cumulants~(\ref{eq:cumulants}) of a single random walker step, the first few non-vanishing being
\begin{equation*}
\kappa_2=\frac13, \quad \kappa_4=-\frac2{15}, \quad \kappa_6=\frac{16}{63}, \quad \kappa_8=-\frac{16}{15},\,\quad \ldots
\end{equation*}
or, as a general expression
\begin{equation}
\kappa_n= \frac{2^n B_n}{n} \quad\quad \mbox{for} \quad n\geq 2,
\end{equation}
where $ B_n $ denotes the $ n $-th \emph{Bernoulli number}.
In order to prove the central limit theorem, we consider the sum over $ N $ subsequent steps of the random walker
\begin{equation*}
Y = X_0+X_1+ ... + X_{N-1},
\end{equation*}
where the random variables $ X_0, X_1, \ldots $ denote the single moves. Since all $ X_j $ are drawn from the same distribution, the index $ j $ can be omitted and the cumulant ratio after $ N $  steps, $ R_n^{(N)}:=R_n[Y] $, reads
\begin{eqnarray}
R_n^{(N)}& = \frac{\kappa_n[Y]}{\kappa_2[Y]^{n/2}}
= \frac{\sum_{j=0}^{N-1}\kappa_n[X_j]}{\left(\sum_{j=0}^{N-1}\kappa_2[X_j]\right)^{n/2}}
= \frac{\sum_{j=0}^{N-1}\kappa_n}{\left(\sum_{j=0}^{N-1}\kappa_2\right)^{n/2}}\nonumber\\
&= \frac{N\kappa_n}{(N\kappa_2)^{n/2}}
= \frac{R_n}{N^{n/2-1}}
= \left\{
\begin{array}{cl}
1 &\quad\mbox{for}\quad n=2, \\
0 &\quad\mbox{for}\quad n>2\quad\mbox{and}\quad N\rightarrow\infty.	
\end{array}\right.
\end{eqnarray}
Hence, in the limit $ N\rightarrow\infty $, the sum $Y$ exhibits a symmetric Gaussian distribution with variance $ \sigma_Y^2=N \kappa_2 $, reproducing the central limit theorem for an ordinary random walk with bounded displacements.

\section{Self-inflating random walk}
Let us now turn to a random walk on a self-inflating background. To this end we first rewrite the dynamical equation of an ordinary random walk (\ref{eq:OrdinaryRW}) in the form
\begin{equation}
\frac{\d}{\d t} x(t) \;=\; \sum_{j=0}^{N-1} \, r_j \, \delta(t-t_j)\,.
\end{equation}
where the $ t_j $ denote the time instances when the jumps take place. In this equation the effect of self-expansion can be incorporated by simply including the term on the r.h.s. of Eq.~(\ref{eq:selfinflation}), i.e.
\begin{equation}
\frac{\d}{\d t} x(t) \;=\; H(t)x(t) + \sum_{j=0}^{N-1} \, r_j \, \delta(t-t_j)\,.
\label{eq:equation_of_motion}
\end{equation}
Introducing the rescaled position $s(t)={x(t)}/{a(t)}$, one can easily show that the Hubble parameter drops out, leading to
\begin{equation}
\frac{\d}{\d t} s(t) \;=\;  \sum_{j=0}^{N-1} \, \frac{r_j}{a(t_j)} \, \delta(t-t_j)\,.
\end{equation}
Hence, the rescaled position of the random walker at time $t_N$ will be given by the sum
\begin{equation}
s_N \;:=\; s(t_N) \;=\; \sum_{j=0}^{N-1} r_j\, w_j
\label{eq:final_rescaled_position}
\end{equation}
of statistically independent random numbers $r_j$ weighted by a factor
\begin{equation}
	w_j:=w(t_j)=\frac{1}{a(t_j)}.
\end{equation} 
On an inflating space, where $a(t)$ increases with time, this means that the weights decrease from move to move. In fact, as we saw in Fig.~\ref{fig:rw}, the stretch is maximal for early moves. 

\begin{figure}
	\centering\includegraphics[width=0.65\textwidth]{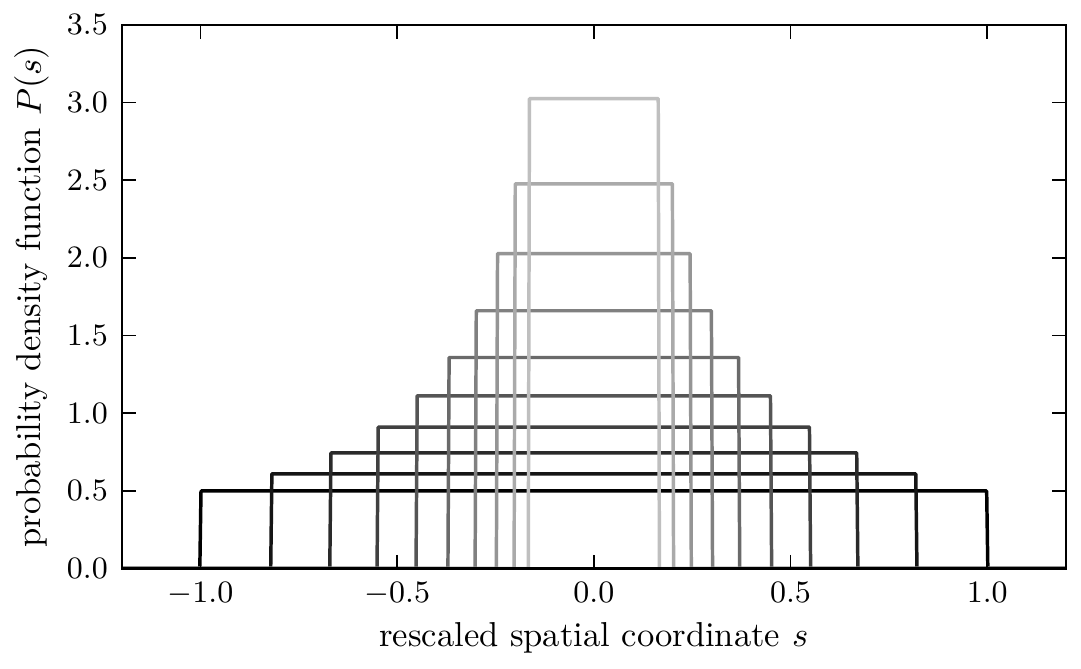}
	\caption{\textbf{Weighted probability density.} Starting from the symmetric uniform distribution of Eq.~\eqref{eq:uniform_distribution}, the succeeding distributions become narrower with time (lighter colors) in the rescaled frame of reference. Drawn are the first ten steps, with constant time steps $ \Delta t=1 $. }
	\label{fig:distributions}
\end{figure}

\subsection{Cumulants of a self-inflating random walk}

As pointed out in the preceding section, the random walk on an inflating background is equivalent to a random walk on a static background with weighted steps. Following Eqs.~\eqref{eq:uniform_distribution} and \eqref{eq:final_rescaled_position}, the probability density of a single step $ X_j $ in the rescaled frame of reference now reads
\begin{equation}
	P_{X_j}(s)=\mathcal{U}_{(-w_j,w_j)}(s) = \left\{
	\begin{array}{cl}
		\frac{1}{2w_j}&\quad\mbox{for}\quad -w_j<s<w_j,\\
		0 &\quad\mbox{otherwise}.
	\end{array}\right.
\end{equation}
Fig.~\ref{fig:distributions} depicts the time evolution of $ P_{X_j} $ schematically. 
It turns out that the random variable $ X_j $, which determines the $ j $-th step of the random walker, is obviously given by $ X_j=w_j X_0 $, whereat $ X_0 $ denotes the random variable of the initial step ($ w_0=1 $). However, in order to avoid any misunderstandings, it is important to notice that for the particular realizations $ X_0=r_0 $ and $ X_j =r_j $ in general $ r_j\neq w_j r_0 $ holds since the $ r_j $ are independently generated random numbers. We can now easily calculate the $ n $-th cumulant of $ X_j $ using the homogeneity property
\begin{equation}
\kappa_n[X_j] = \kappa_n[w_j X_0] = w_j^n \kappa_n[X_0] = w_j^n\, \frac{2^n\,B_n}{n}.
\end{equation}
Furthermore, the cumulants after $ N $ subsequent steps,
\begin{equation}
Y=X_0+X_1+...+X_{N-1},
\label{eq:nonstandardized_sum}
\end{equation}
can be obtained by simply summing up all single cumulants
\begin{equation}
\kappa_n^{(N)}:=\kappa_n[Y]= \sum_{j=0}^{N-1}\kappa_n[X_j].
\label{eq:cumulants(N)}
\end{equation}
If we calculate the cumulant ratio \eqref{eq:cumulant_ratio_definition} of the $ j $-th step which we label $ R_n[X_j] $, the weights cancel, thus giving
\begin{equation}
R_n[X_j] = R_n[X_0] = \frac{\kappa_n}{\kappa_2^{n/2}} = \frac{(2\sqrt{3})^n\,B_n}{n}.
\end{equation}
Finally, the cumulant ratio after $ N $ steps is given by the following expression
\begin{equation}
R_n^{(N)} = \frac{\kappa_n^{(N)}}{\left(\kappa_2^{(N)}\right)^{n/2}} =  \frac{\kappa_n}{\kappa_2^{n/2}}\; \frac{\sum_{j=0}^{N-1}w_j^n}{\left(\sum_{j=0}^{N-1}w_j^2\right)^{n/2}}.
\label{eq:cumulant_ratio(N)}
\end{equation}

\subsection{Exponentially expanding support}

We now consider the special case of an exponential expansion $ a(t)=a(t_0)\,\mathrm{e}^{\,\mu (t-t_0)} $ where $ \mu $ controls the expansion time scale. This corresponds to a choice of $ H(t)=\mu=\mathrm{const} $ for the Hubble parameter. We calculate the sum \eqref{eq:nonstandardized_sum} of $ N $ subsequent random walker steps. For convenience, we set $ t_0=0 $, $ a(t_0)=1 $ and use a time step of $ \Delta t=1 $, thus giving weight factors of $ w_j =1/\mathrm{e}^{\,\mu j}  $. The cumulants \eqref{eq:cumulants(N)} then read
\begin{equation}
	\kappa_n^{(N)}=\frac{2^n\,B_n}{n}\sum_{j=0}^{N-1}\frac{1}{\mathrm{e}^{n\mu j}}  =\kappa_n\; \frac{1-\mathrm{e}^{-n\mu N}}{1-\mathrm{e}^{-n\mu}},
\end{equation}
which, as $ N\rightarrow \infty$, results in
\begin{equation}
	\kappa_n^\infty = \frac{\kappa_n}{1-\mathrm{e}^{-n\mu}} = \mathrm{const.} \neq 0 \quad \forall \mu> 0 \quad\mbox{and}\quad \forall n \quad\mbox{even}.
\end{equation}
Consequently, also the cumulant ratio $R_n^\infty\neq0$ and the central limit theorem fails in this case, indicating that on an exponentially inflating support, the random walk loses its characteristic behavior. Moreover, for large $ \mu $, the above expression becomes $ \kappa_n^\infty =\kappa_n $ which is nothing else than the uniform distribution of the first step. This is a reasonable result, since for very large $ \mu $ the footprint of the initial step dominates all remaining moves.

\subsection{Algebraically expanding support}
\label{sec:Algebraically_expanding_support}
\begin{figure}[t]
	\centering\includegraphics[width=0.49\textwidth]{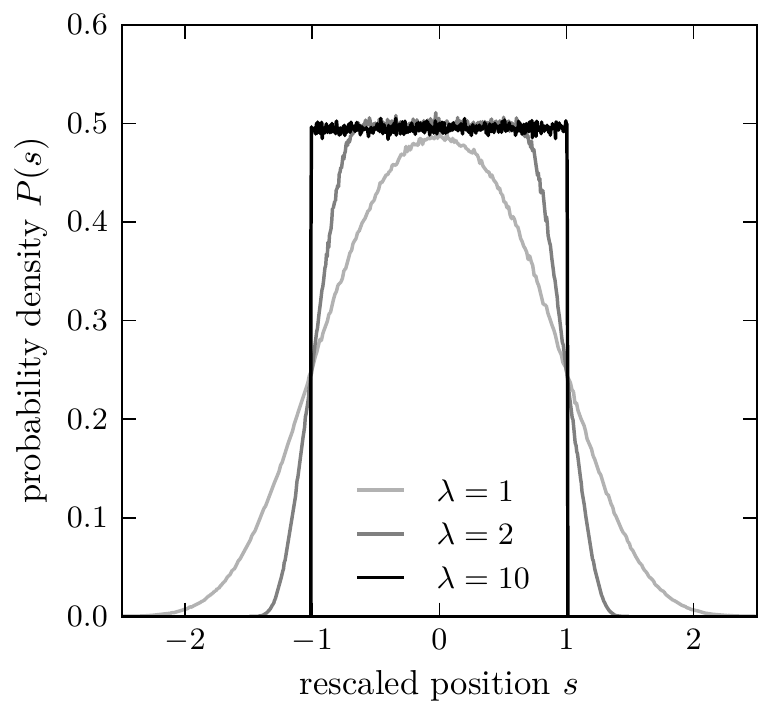}
	\centering\includegraphics[width=0.49\textwidth]{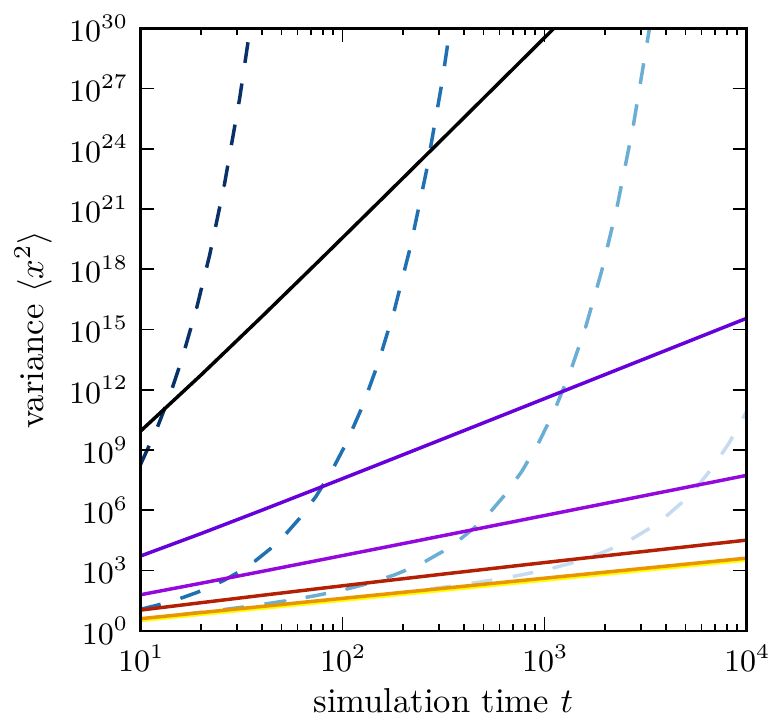}
	\caption{\textbf{Numerical results.} Left panel: Resulting distributions on algebraically expanding backgrounds for different values of the expansion control parameter, binned from $ 10^7 $ Monte Carlo runs \`{a} $ 10^3 $ steps. Right panel: Mean quadratic distance from the origin over time for random walks on differently strong expanding backgrounds. Each data point was averaged over $ 10^6 $ Monte Carlo simulations. The solid straight lines denote algebraic expansion, with $ \lambda=0,\,0.1\,,0.5\,,1.0\,,2.0\,,5.0 $ in ascending order. The dashed lines denote exponential expansion with $ \mu=0.001,\, 0.01,\, 0.1,\, 1$, again in ascending order.\vspace{3mm} }
	\label{fig:numerical_results}
\end{figure}

As exponential background growth is obviously too fast to conserve the normal distribution of a random walker, we study now the behavior in the case of an algebraical expansion, i.e. the expansion is directed by a power law. According to Equation~\eqref{eq:scalefactor}, the expansion rate $H(t)=\lambda/t$ leads to a scale factor $a(t)=(t/t_0)^\lambda$. Hence the cumulants~\eqref{eq:cumulants(N)} are given by
\begin{equation}
	\kappa_n^{(N)}=\frac{2^n\,B_n}{n}\sum_{j=1}^{N}\frac{1}{j^{n \lambda}} =\kappa_n \sum_{j=1}^{N}\frac{1}{j^{n \lambda}},
	\label{eq:cumulant_algebraic(N)}
\end{equation} 
using time steps of $\Delta t=1$ and setting $ t_0=1 $. In order to determine for which choices of $ \lambda $ the central limit theorem is still valid, we consider the cumulant ratio \eqref{eq:cumulant_ratio(N)}. It reads
\begin{equation}
	R_n^{\infty}=\frac{\kappa_n}{\kappa_2^{n/2}}\frac{\sum_{j=1}^{\infty}\frac{1}{j^{n \lambda }}}{\left(\sum_{j=1}^{\infty}\frac{1}{j^{2\lambda}}\right)^{n/2}}.
\end{equation}
By definition, $R_2^{\infty}=1$ holds for the second cumulant ratio, whereas for $n>2$ we have to consider three different cases:
\begin{itemize}
 
\item For $0\leq\lambda\leq1/n$ the denominator diverges faster than the numerator so that the cumulant ratios $R_n^\infty$ vanish. This can be seen if one compares both series, since for each two corresponding summands the inequality  $\frac{1}{j^{2\lambda}}>\frac{1}{j^{n\lambda}}$ is fulfilled. 
\item In the range $1/n<\lambda\leq1/2$ the numerator converges whereas the denominator diverges, thus $R_n^{\infty}\rightarrow0$ still holds. 
\item For $\lambda>1/2$ one can write

\begin{equation}
	R_n^{\infty}=\frac{\kappa_n}{\kappa_2^{n/2}}\frac{\zeta(n \lambda)}{\zeta(2\lambda)^{n/2}},
\end{equation}
where $\zeta(x)=\sum_{m=1}^{\infty}m^{-x}$ denotes a representation of the Riemann zeta function. It is monotonously decreasing for $x\in(1,\infty)$ with 
\begin{equation}
	\lim\limits_{x\rightarrow 1}\zeta(x)=\infty \quad\mbox{and}\quad\lim\limits_{x\rightarrow \infty}\zeta(x)=1.
	\label{eq:limitzeta}
\end{equation}
Therefore, it is easy to see that $0< R_n^{\infty}< \frac{\kappa_n}{\kappa_2^{n/2}}$ for $n>2$ and $\lambda>1/2$.
\end{itemize}
Summing up, we have
\begin{equation}
	R_n^{\infty}=\left\{\begin{array}{cl}
		0 & \quad\mbox{ if}\quad 0\leq\lambda\leq\frac{1}{2} \\
		\frac{\kappa_n}{\kappa_2^{n/2}}\frac{\zeta(n \lambda)}{\zeta(2\lambda)^{n/2}}\neq 0 & \quad\mbox{ if}\quad \lambda>\frac{1}{2}
	\end{array}\right.\,\quad\quad\mathrm{for\ even}\ n>2.
\end{equation}
For this reason, we conclude that the central limit theorem still holds if $\lambda\leq\frac{1}{2}$, but fails for stronger background growth. Apparently, the expansion process prevails over the random walk  in the latter case, since the underlying support expands faster than the length scale $\xi\sim t^{1/2}$ of the random walk itself. 
Furthermore, we see from Eqs.~\eqref{eq:cumulant_algebraic(N)} and \eqref{eq:limitzeta}, that in the limit of large $ \lambda $
\begin{equation}
\lim\limits_{\lambda\rightarrow\infty}\kappa_n^\infty =\lim\limits_{\lambda\rightarrow\infty}\kappa_n \sum_{j=1}^\infty\frac{1}{j^{n \lambda}} = \kappa_n\lim\limits_{\lambda\rightarrow\infty}\zeta(n\lambda)=\kappa_n
\end{equation} 
holds. Consequently, for very strong algebraic expansion the distribution becomes again uniform, analogously to the exponential case. This behavior was confirmed by numerical simulations (see left panel of Fig.~\ref{fig:numerical_results}), where the shape of the distribution differs more and more from a Gaussian as $ \lambda $ is increased. Furthermore, the right panel of Fig.~\ref{fig:numerical_results} shows the mean quadratic distance from the origin (i.e. the correlation length squared) for different choices of the expansion control parameter in both the exponential (dashed lines) and the algebraic case (solid lines). These numerical calculations support our analytical results, since it can be seen that only the solid lines for $ \lambda=0, 0.1 $ and $ 0.5 $ stay parallel, whereas all other curves (especially the exponential ones) diverge faster.\\

\subsection{Dynamical exponent}

The previous results indicate that the overall dynamics of the system can be split into a \emph{diffusion} and an \emph{inflation} component. When interested in the dynamical exponent of the system, we may therefore evaluate it for both sub-processes separately. It is commonly known that for ordinary diffusion the reference length of the system $ \xi $ scales as $ \xi \sim t^{1/z} $ with a dynamical exponent of $ z_{\mathrm{Dif}}=2 $. For the inflation however, the dynamical exponent is simply given by $ z_\mathrm{Inf} = 1/\lambda $ for a power-law driven expansion, since the length scale of the system scales as $ \xi \sim a(t)= t^{\lambda} $ in this case.

\begin{figure}[h]
 	\centering\includegraphics[width=0.49\textwidth]{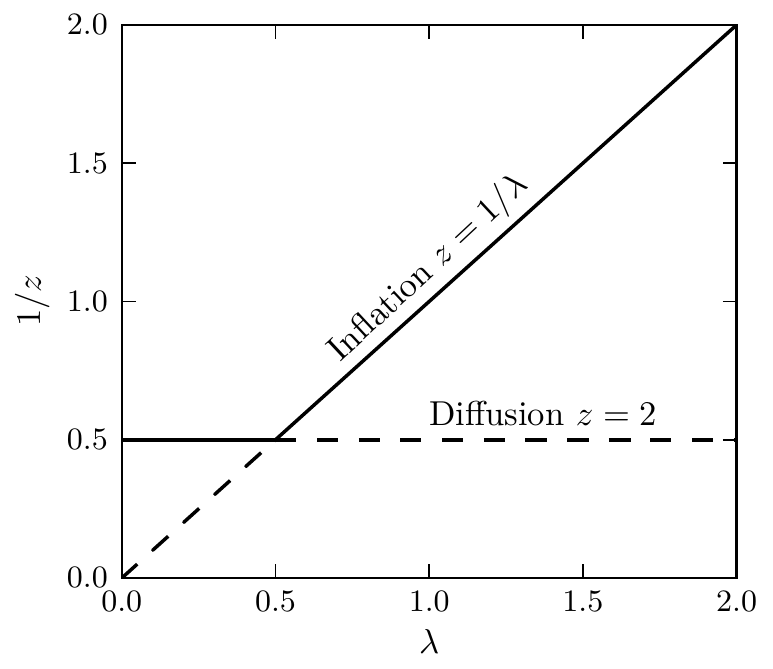}
 	\caption{\textbf{Dynamical exponent.} Shown is the inverse dynamical exponent as a function of the expansion parameter $\lambda$. For $\lambda>1/2$, the inflation starts to dominate the dynamics and the system shows super-diffusive behavior.}
 	\label{fig:dynamical-exponent}
 \end{figure}

In Fig.~\ref{fig:dynamical-exponent} the inverse dynamical exponent is plotted against $ \lambda $ for both competing processes. There, the solid curve indicates the effective dynamical exponent of the full system, whereas the dashed lines denote the inferior processes. Altogether, we found that for the algebraically expanding support the dynamical exponent is given by
\begin{equation}
	z = \min\left(z_\mathrm{Dif},z_\mathrm{Inf}\right)= \min\left(2,\lambda^{-1}\right) 
	= \left\{\begin{array}{cl}
	2\quad &\mbox{for}\quad \lambda\leq \frac{1}{2}, \\
	\frac{1}{\lambda}\quad &\mbox{for} \quad\lambda> \frac{1}{2}.
	\end{array}\right.\,
\end{equation}

Finally, note that for exponential growth the dynamical exponent is meaningless since the inflation as the dominating sub-processes behaves non-algebraic in this case. 
 
\section{Conclusion}

In the present work we have studied the behavior of a random walk on a self-inflating background which expands isotropically and homogeneously with time. The expansion is described by the Hubble parameter $H(t)=\dot{a}(t)/a(t)$, where $a(t)$ is the associated scale factor. 

Two cases have been considered, namely, an exponential inflation $H(t)=\mu=\mathrm{const}$ and an expansion with power-law characteristics where $H(t) = \lambda/t$.  For exponential inflation, it turns out that the key properties of the random walk, most notably its scale invariance, is destroyed after some time since the random walker cannot propagate as fast as the background expands. For a power-law driven expansion, however, the situation depends on the value of the exponent $\lambda$. For $0<\lambda \leq \frac12$ the central limit theorem still holds so that a Gaussian distribution is obtained. For $\lambda> \frac12$, however, the expansion is so fast that higher cumulant ratios do not vanish. Here, the process is initially governed by diffusion with the dynamical exponent $z=2$, crossing over to an accelerated expansion with $z=1/\lambda < 2$ where the rescaled probability density is essentially frozen. We have only considered the case of a random walk in one dimension, since a generalization to a multi-dimensional setting (i.e. isotropic random walk as well as isotropic expansion) is straightforward as our results can be applied to each dimension individually.

This study has been focused on the random walk as the simplest example of a self-similar stochastic process. It would be interesting to see how other self-similar stochastic processes such as systems at the critical point of a phase transition behave on a self-inflating background. 


\end{document}